\documentclass[sigconf,nonacm]{acmart} 
\copyrightyear{2026}
\acmYear{2026}
\setcopyright{cc}
\setcctype{by-nc-nd}
\acmConference[LLM4Code '26]{3rd International Workshop on Large Language Models For Code }{April 12--18, 2026}{Rio de Janeiro, Brazil}
\acmBooktitle{3rd International Workshop on Large Language Models For Code (LLM4Code '26), April 12--18, 2026, Rio de Janeiro, Brazil}

\usepackage{graphicx} 

\title{Achieving Productivity Gains with AI-based IDE features: A Journey at Google}
\titlenote{Accepted for publication at the 3rd International Workshop on Large Language Models For Code (LLM4Code '26).} 

\author{Maxim Tabachnyk}
\affiliation{%
  \institution{Google}
  \city{Munich}
  \country{Germany}
}
\email{tabachnyk@google.com}

\author{Xu Shu}
\affiliation{%
  \institution{Google}
  \city{New York City}
  \country{USA}
}
\email{xushu@google.com}

\author{Alexander Frömmgen}
\affiliation{%
  \institution{Google}
  \city{Munich}
  \country{Germany}
}
\email{froemmgen@google.com}

\author{Pavel Sychev}
\affiliation{%
  \institution{Google}
  \city{Munich}
  \country{Germany}
}
\email{luckygeck@google.com}

\author{Vahid Meimand}
\affiliation{%
  \institution{Google}
  \city{Washington}
  \country{USA}
}
\email{vahidzm@google.com}

\author{Ilia Krets}
\affiliation{%
  \institution{Google}
  \city{Munich}
  \country{Germany}
}
\email{iliakrets@google.com}

\author{Stanislav Pyatykh}
\affiliation{%
  \institution{Google}
  \city{Munich}
  \country{Germany}
}
\email{pyatykh@google.com}

\author{Abner Araujo}
\affiliation{%
  \institution{Google}
  \city{Munich}
  \country{Germany}
}
\email{abneraraujo@google.com}

\author{Kristóf Molnár}
\authornote{Work carried out while at Google. At the time of publication working for Databricks.}
\affiliation{%
  \institution{Google}
  \city{Munich}
  \country{Germany}
}
\email{krimol@google.com}

\author{Satish Chandra}
\authornote{Work carried out while at Google. At the time of publication working for Meta.}
\affiliation{%
  \institution{Google}
  \city{Sunnyvale}
  \country{USA}
}
\email{schandra@acm.org}

\setcopyright{none}
\makeatletter
\renewcommand\@formatdoi[1]{\ignorespaces}
\makeatother
\renewcommand\footnotetextcopyrightpermission[1]{} 
\begin{document}

\begin{abstract}
    We discuss Google's journey in developing and refining two internal AI-based IDE features: code completion and natural-language-driven code transformation (Transform Code).  We address challenges in latency, user experience and suggestion quality, all backed by rigorous experimentation. 
    The article serves as an example of how to refine AI developer tools across the user interface, backend, and model layers,  to deliver tangible productivity improvements in an enterprise setting.
\end{abstract}

\maketitle

\section{Introduction}

Modern IDEs come with a host of AI-based features served by large-language models (LLMs).  At first blush, these features might appear to be straightforward applications of LLMs, but the path from building these features in to getting them to deliver measurable productivity gains when deployed at scale---arguably the purpose of AI enablement in the first place---is non-trivial.

\begin{figure}[h]
\includegraphics[scale=0.2]{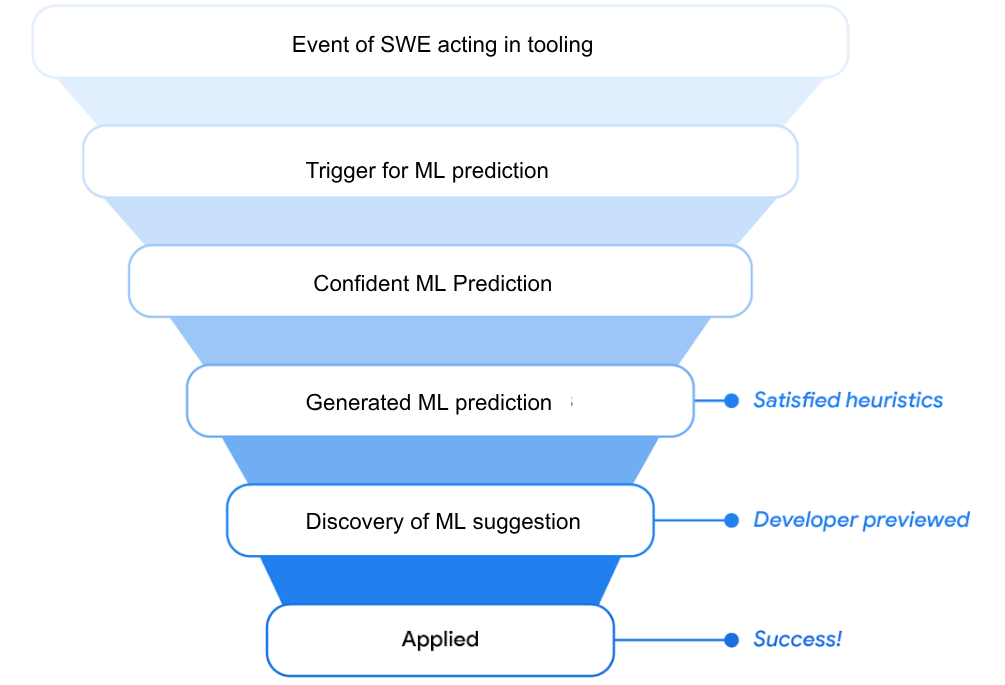}
\caption{An opportunity funnel starting from the software engineer's (SWE) actions down to actual application of ML-based suggestions. Opportunities are lost if the model prediction is not confident enough, the model doesn’t respond or responds too late, the prediction is subpar, the user doesn’t notice the prediction, and so on.\label{fig:funnel}}
\end{figure}

An AI-based suggestion in an IDE has not only has to be high quality, but it also needs to be discoverable in the IDE, respond with low latency, and generally be easy-to-review, to attract high engagement from the users.  Fig~\ref{fig:funnel} presents this as a ``funnel diagram'' of opportunities lost along the way. To get this engagement and consequently accomplish productivity gains, we need to address all of these challenges.  

We describe our journey down this path with two AI-based features: code completion and code transformation. Both are features in Google’s internal IDE and are mature products, measurably adding value to thousands of daily active users. We have developed additional AI-based features as well \cite{ChandraTabachnyk2024}, but the two that we discuss nicely illustrate the pattern of user-experience considerations and data-driven iterations.

\textbf{AI-powered code completion} predicts how to continue the code after the current cursor position with a few words or up to multiple lines. It anticipates the user's coding intent based on the IDE context and provides real-time suggestions of what the user is likely to type next. It has become an integral feature of modern IDEs, both at Google and across the industry. \cite{murali2024aiassistedcodeauthoringscale,xu2023}

\begin{figure}
\includegraphics[scale=0.25,trim={0 0 6cm 0},clip]{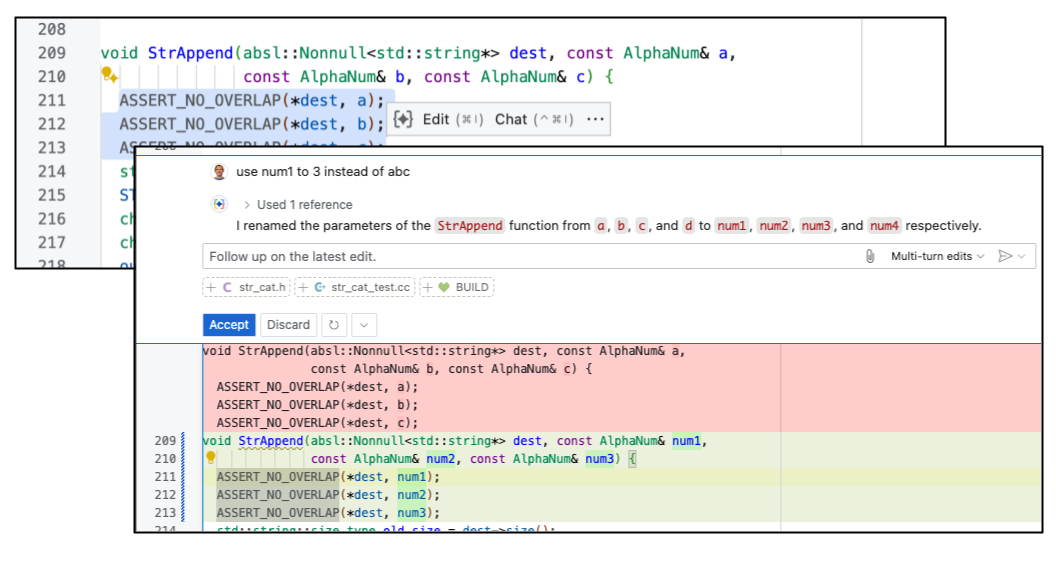}
\caption{Transform Code in action. The user selects from code (background), and can choose to enter an edit prompt. The foreground shows the AI's response to the user prompt. \label{fig:tc}}
\end{figure}

\textbf{Transform Code} is a feature for natural-language-driven code editing. See Fig~\ref{fig:tc}. The developer triggers the feature manually (e.g. by clicking on "edit" popup; background inset) and specifies the intended (small) code generation or edit in a natural language prompt (here, \textit{use num1 to 3 instead of abc}, as shown in the foreground inset).  The foreground also shows shows the edit suggestion from Transform Code in response to the user's prompt. 
Examples of prompts include: \textit{log undoStack length}, \textit{add details about the params}, \textit{add tests}, or simply \textit{fix}. Transform Code is intended for small changes as opposed to complete implementation. Such an "inline edit" feature is also becoming standard in modern AI-enabled IDEs (e.g. Cursor\cite{cursor:website}).

\paragraph*{Outline and Contribution}

The article describes (Sections~\ref{sec:cc},~\ref{sec:tc}) the progressive improvements that we made, all carefully backed up with \textbf{data-driven decisions from online experimentation} rolled out to our internal developer population. This is in contrast with relying on our intuitions or only offline tests.

These AI features have led to \textbf{measured productivity gains} at Google. The article also describes (Sec~\ref{sec:productivity}) how we measure the productivity improvement that we see in deployment at scale.  We conclude the article with an outlook for the future.

While there is growing literature on building  AI-based developer tools atop LLMs for various purposes, the article serves as an exemplar of how to \textit{refine} such tools, with focus on user experience and online experimentation.  We believe such an overall approach to practical productivity tools is universally applicable.  

\section{AI-based code completion}
\label{sec:cc}
\subsection{Basic model} 

The AI training task in code completion is a fill-in-the-middle (FIM) completion of code after the cursor, where the input contains code from the current file and is augmented with additional context.   Our supervised fine-tuning (SFT) data relies on the editing history of Google engineers at keystroke-level granularity, keeping us in-distribution to production traffic at training time. 

\subsection{Challenges}

We previously reported on our initial AI code completion efforts on smaller encoder-decoder models (0.5B parameters) in \cite{TabachnykNikolov2022}. Larger models can increase the quality of suggestions and be more helpful in complex scenarios. However, a larger model comes with new challenges, notably in \textbf{latency and computational costs}.  We applied the following techniques to be able to scale the model size:
\begin{itemize}
\item Implement various caching and reuse strategies (described in Sec~\ref{sec:adaptive}.)
\item Use the smallest possible model with Speculative Decoding \cite{10.5555/3618408.3619203} and compensate for the drop in quality by doing SFT with high-quality data.
\item Limit tail latency and cost by constraining the number of input (8K) and output (128) tokens, as well as compact format for describing input context and predictions.
\end{itemize}

A second challenge is to boost \textbf{suggestion quality} by leveraging the larger context window provided by modern LLMs; see Section~\ref{sec:context}.

Metric-wise, our goal was to increase the feature’s impact, measured as the fraction of code written by ML (FCML), while keeping acceptance rate steady, which we treat as a proxy for quality and user annoyance. \textbf{FCML} is defined as the number of accepted characters from AI-based suggestions divided by the sum of manually typed characters, copy-pasted characters (under 1000 character length and excluding full-file pastes) and accepted characters from AI-based suggestions. \textbf{Acceptance rate} is defined as the number of accepted suggestions divided by the sum of accepted and rejected suggestions, where we only count rejected suggestions if they were visible for at least 750 ms to the user.

\subsection{Adaptive Caching to mitigate latency}
\label{sec:adaptive}
Conventional remote procedure call (RPC) \cite{google_cloud_protorpc_2025} caching and de-duplication reduces the volume of requests (by 15\%), saving compute serving resources. However, during the rollout of the new stack we discovered the following issues that need to be addressed:
\begin{itemize}
\item Fast forward typing: as the model input changes with every keystroke, we are not fast enough to present suggestions even if the user is typing what the model might have predicted.

\item Local stability of consecutive suggestions: if for \texttt{B()} we suggest \texttt{Build()}, users expect that for \texttt{Bu()} we also suggest \texttt{Build()}. In practice, due to tokenization artifacts, the next prediction can be different, e.g. \texttt{Bundle()}. 

\item Fast typing of a single user leads to bursts of compute usage, causing stalls for other users due to the exhaustion of the shared resource.
\end{itemize}

To address these issues, we implemented "adapting cache" to manage in-flight and finished requests (“streak”), trying to reuse their results. When a completion request arrives:
\begin{itemize}
\item Existing streak responses are adapted if possible, e.g. reply \texttt{ild} to \texttt{Bu()} if  \texttt{Build()} is in the cache. We pick the oldest matching response to achieve local stability.
\item Otherwise, a model request is scheduled and queued if there are already two in-flight requests. 
\item The system waits for either of (a) request cancellation (e.g. user typed ahead). We remove the queued request if it has not started processing to save resources; or (b) model responds to some inflight request, the system adapts this response to waiting requests, and replies where possible. The next queued request gets executed.
\end{itemize}

\subsection{Providing Useful Context}
\label{sec:context}
We consider prompt building as a packing problem, maximizing the number of relevant code snippets within a prompt size budget. This is accomplished with two steps:
\begin{enumerate}
\item Selection and ranking of relevant snippets
\item Rendering those snippets into the prompt in a concise and understandable way. 
\end{enumerate}

Step 1: We track edits done in all open files, ranked by their recency. Additionally, we scan the opened files with a sliding window to find textual matches with the code around the cursor. They are ordered by the number of matched words and are always ranked lower than the edits. We use case-insensitive, word-level matching in order to capture code entity references in comments and associate function names with their corresponding variables and class members, e.g. \texttt{ComputeAnnualBalance()} $\leftrightarrow$ \texttt{annual\_balance}.

\begin{figure}
\includegraphics[scale=0.3]{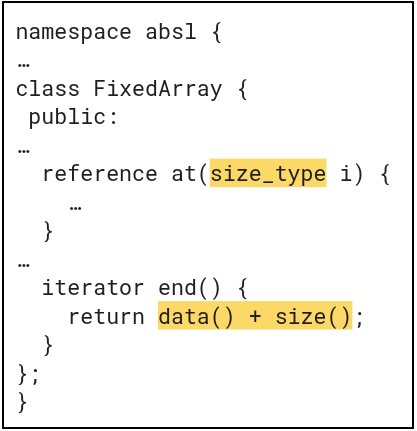}
\caption{Example of syntax-preserving file rendering for C++. Relevant snippets are marked in yellow. Pruned pieces are replaced with a single-token marker "…". \label{fig:cc1}}
\end{figure}

Step 2:  Each relevant snippet is rendered into the prompt together with all scopes enclosing it. For example, a snippet containing class members is rendered together with the respective class header, and a snippet within a function body is rendered together with the respective function signature, even if the latter is far away in the file. An example is shown in Figure \ref{fig:cc1}. This ensures that the model gets each snippet in the proper context, which is essential to utilize it correctly.

\subsection{Results}

The caching approach led to 35\% cache hit rate, end-to-end p50 (median) latency  reduced by 9\%, p90 (the top 10 percentile) by 2\%. Acceptance rate increased by 17\%. The unusual combination of showing more suggestions, and having a higher overall acceptance rate led to a large relative increase of 41\% in FCML. 

As for the context optimization, in an A/B experiment (random assignment to control and treatment from Google’s SWE population, 2.5K+ users per group, duration 2 weeks), the acceptance rate increased by 5\%, and FCML increased by 11\%. Since the size of the prompt increased, the median and the p90 latency went up by 46\% and 10\% respectively. This resulted in the amount of shown suggestions going down by 5\%.

Cumulatively, with the above improvements, in March 2025 over 4 weeks, we achieved the success metrics presented below:

\begin{tabular}{lc}
\hline
  Fraction of code written by ML (FCML) & 28.7\% \\
FCML excluding paste in denominator & 70.6\% \\
Acceptance rate  & 45\% \\
Average characters added per accept & 62 \\
\hline\\
\end{tabular}

\section{Transform Code}
\label{sec:tc}
\subsection{Basic model}
We use Gemini models that are additionally trained on Google internal code similar to \cite{10.1145/3639477.3639746}. The additional training teaches the model relevant knowledge about internal code and improves its capability to handle our input and output formats. The input format consists of the relevant files,  a short repetition of the content around the cursor (or selected code) together with explicit markers for the beginning and the end of the selected code, and the natural language prompt of the user. The output format uses a compact edit representation inspired by the Unified Diff Format. \cite{gnu:diffutils:unified} The output format relies on three anchor lines, i.e., lines that are repeated from the input file without any modification, to find the relevant code and apply the edit. 

\subsection{Challenges}

Applying LLMs for natural-language driven code editing in the IDE came with some challenges, among which:

\begin{itemize}
\item \textbf{Discoverability} to guide users to the entry point in relevant situations (Sec~\ref{sec:discover}.)
\item \textbf{Easy-to-review} suggestions of complex code edit suggestions (many lines, many locations across one or multiple files) (Sec~\ref{sec:render}.)
\item High-enough \textbf{suggestion quality} for potential multi-turn prompts and contexts common in the IDE containing in-progress code (Sec~\ref{sec:distribution}.)

\end{itemize}

\subsection{Discoverable entry points}
\label{sec:discover}
Since the feature needs to be actively triggered by the user, a discoverable entry point is critical for scaling usage when the feature quality reaches a sufficiently high level. The challenge with adapting the entry point UX is to find a balance of showing it often enough in situations when the feature is relevant, versus not distracting the user too often.

In qualitative UX studies, we found that one common situation where Transform Code becomes useful is right after they select a section of their code. See Fig~\ref{fig:tc}, the background inset. We experimented with showing a floating button next to the code selection that allows triggering the feature either with a mouse click, while also exposing the keyboard shortcut, so users can learn it.

\subsection{Improved edit rendering for accelerating suggestion review}
\label{sec:render}

Aside from the time spent on writing prompts, reviewing suggestions is a key time cost when using the feature. Our hypothesis was that we can accelerate the suggestion review by improving the code diff visualization (see Figure \ref{fig:unwanted}):

\begin{itemize}
\item Minimizing the number of decorated lines and highlighted words: now focusing diff rendering on specific minimal set of truly edited lines vs. before blocks of edited code where some lines could be without an edit 
\item Differentiating moved lines from changed lines: now tagging moved lines explicitly as "moved" (CodeLens VSCode UI) in diff vs. before showing deleted and created code blocks.
\end{itemize}

\subsection{User rewrites to bridge the distribution gap}
\label{sec:distribution}
\begin{figure}

\includegraphics[scale=0.15]{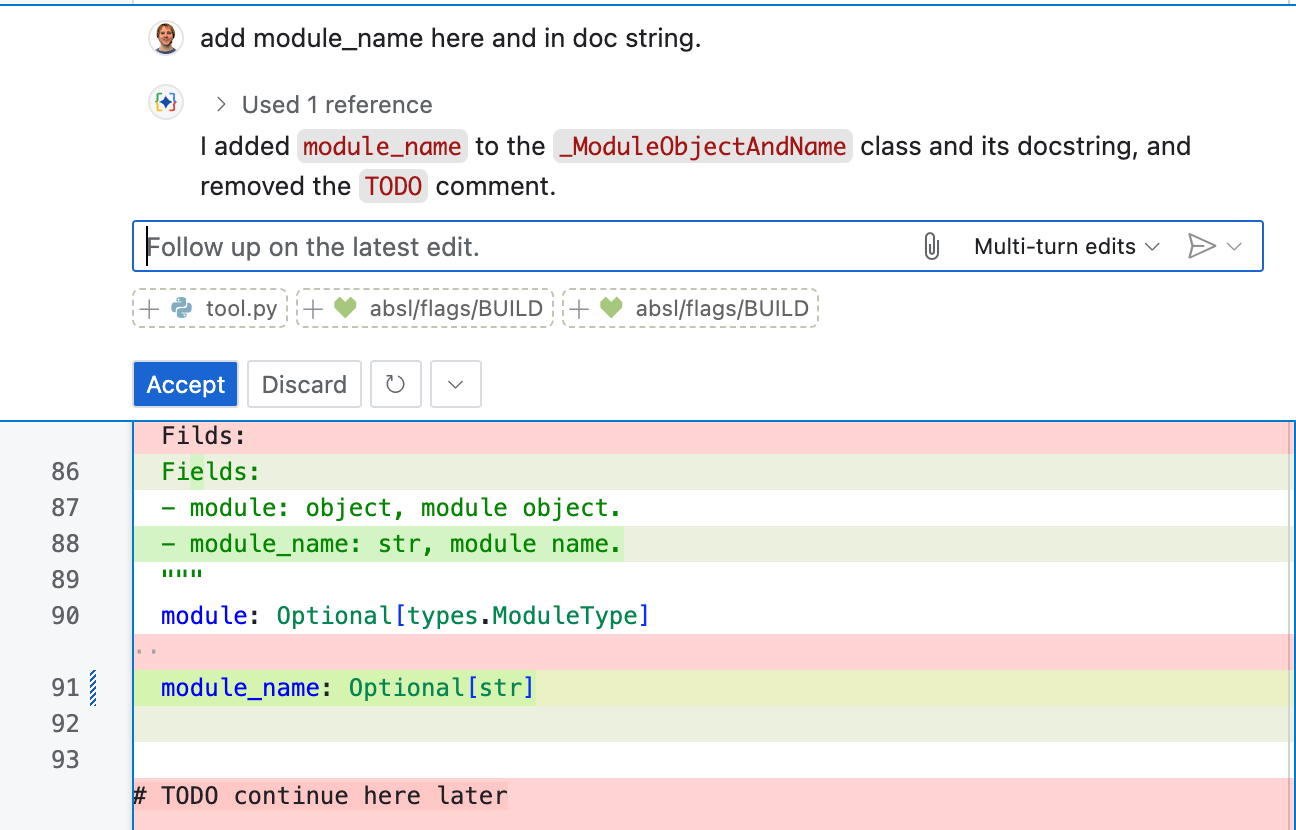}
\caption{Examples of undesired edit suggestions that got resolved by training on high quality user rewrites. In this example, the model fixes an additional typo and removes in-progress code snippets. \label{fig:unwanted}}
\end{figure}

We found that the model supporting Transform Code struggled in three related dimensions: instruction following, introduction of unrelated edits and removal of work in-progress code. We attribute this to the distribution mismatch of “high quality code” training data targets (submitted reviewed code) and the work-in-progress nature of the code in the IDE. Fig~\ref{fig:unwanted} illustrates these points.

In order to improve the model, we collected rewrites from day-to-day usage of volunteer Google engineers in the IDE, meaning that after rejecting a Transform Code suggestion, the user could record a manual edit. An extra, manual curation step was needed to further improve the data quality, particularly to remove examples where the user recorded unrelated edits, or when the relevant context was lacking.
We then included O(100) of such high quality examples for SFT as part of the post training.

Incidentally, in order to cater to the expectation of multi-turn conversational support, we use the same mechanism of curated SFT examples.  We created ~200 highly curated conversational code edit examples that combine edits and text responses in the target. We found that mixing these into the post training teaches the model to combine text responses with edits. We mixed additional examples with a conversation history into the training mixture and found them sufficient to teach the model multiturn capabilities.

\subsection{Results}

\paragraph*{Discoverability} We conducted an A/B experiment---random assignment from Google SWE population to treatment and control, with 10k+ users per group over ~1 month---where some users were exposed to the floating buttons described above, in addition to the keyboard shortcut and menu item available to the control group. The treatment group showed an increase in the number of overall prompts by 40\% with positive user perception on usefulness, and the number of users that send at least one prompt increased by 64\%. The discoverability of the shortcut in the menu increased the number of shortcut triggers by 19\%.

\paragraph*{Edit rendering} An A/B experiment---similarly to the above---for an improved rendering of the suggested edit showed a 7\% decrease in review time, and a 2.2\% increase in acceptance rate. For predictions that resulted in larger diffs (>133 characters), we found a 4.5\% increase in acceptance rate.

\paragraph*{Bridging the Distribution Gap}

Offline evaluation on our curated evaluation dataset---which is a split of the collected rewrites---shows an improvement in the median ChrF score \cite{popovic-2015-chrf} from 84.5 to 88.8, a reduction of edits in unrelated files from 13 to 0.  

Our A/B experiments comparing models with and without supervised fine tuning with ~400 of these rewrites show an acceptance rate improvement from 55\% to 63\%.  Our A/B experiment with the conversational fine tuning shows a small reduction in acceptance rate for this model and prompt format. We have an ongoing evaluation of the benefit of the conversational multiturn capability and allow the user to choose between the models in the meantime.

Here we show our overall feature metrics after all of the above improvements. 

\begin{tabular}{lc}
\hline
Acceptance rate  & 68\%  \\
Average prompts per user per week & 11 \\
Average prompt length in characters & 37 \\
\hline\\
\end{tabular}

We have less than 1s end-to-end latency on average. Although by now features that can tackle significantly more complex tasks are emerging (Sec~\ref{sec:outlook}), we still see a very steep scaling trajectory for this feature as of a few months ago, with a doubling of weekly active users and 3.5x increase of weekly prompts over 3 months.

\section{Measuring Productivity}
\label{sec:productivity}
The feature-specific metrics such as FCML and acceptance rates are “proxy" metrics for productivity. Real productivity shifts are more difficult to measure in practice, and below we discuss some of our approaches.

\subsection{Measuring Impact on Developer Productivity}

Google employs a wealth of qualitative and quantitative data to measure developer productivity. This data informs a comprehensive set of metrics that we utilize to analyze the impact of different tool changes. The metrics discussed here, are a subset which are log-based and have been validated by qualitative studies. The data originates from Google's software engineering toolchain.
\begin{itemize}
\item \textbf{Change List throughput} (CLT): Number of Change Lists (CLs) submitted by developers per month. A CL (or a “diff”) is one self-contained change that has been submitted to version control after undergoing code review.
\item \textbf{Active Coding Time} (ACT) per CL: Active time spent by developers preparing a Change List (hand on keyboard with activity in coding tools related to CL). 
\item \textbf{Mean duration of investigation sessions} (MeanDurInvSess): We define investigation sessions as sessions when developers leave the IDE to go to other internal surfaces that contain relevant information. The period from the first search until they go back to code editing is considered an investigation session.
\end{itemize}

We evaluate the impact of AI-powered features on productivity through causal inference, both online (A/B tests) and offline (observational causal analysis). AI-powered features make their impact a while after they are first launched as they become more mature and usage of them gets settled. Conducting A/B experiments over extended periods to capture these long-term effects is costly and requires a group of users to not access them for prolonged time. Therefore we also use offline observational data to analyze users' interaction with features and the relationship to the productivity metrics. 

Here we discuss an observational causal analysis to estimate the impact of Transform Code (TC) on developers who started using it. We focus on TC since its gradual adoption over time (since it needs proactive user invocation) allowed for this type of study to be done. Notably, this approach would not be possible for auto-triggered code completion. Moreover, some industry studies on the impact of code completion on productivity are already available whereas to our knowledge none for TC-like features is available.

\subsection{Methodology}

We used observational data on developers' interactions with the feature and the Difference-in-Differences (DiD) causal inference framework, specifically the Callaway and Sant’Anna method \cite{CALLAWAY2021200}.   This approach compares the metrics in the adopters versus non-adopters cohorts, adjusting for the difference in the pre-treatment period, i.e. adjusting selection bias.  The method estimates treatment effects for every group (based on when the adoption occurred) and time, and aggregates to the final treatment effect using a weighting schema.  

Our model analyzes the effects of TC adoption on the treatment group by comparing their outcomes after adoption, against a counterfactual scenario that is based on their outcomes before TC adoption, adjusted for changes over time in the control group. 
We also control for the observed confounders to further adjust for the intrinsic difference between the adopters and non-adopters to uncover the causal effect. User-level covariates (role, region, level, tenure) and code-related covariates (e.g. baseline CL size, programming language, CL type) were included in the model.

The treatment group (~36k developers) consisted of users who adopted TC in January-December 2024 and the control group (18k developers) consisted of those who did not adopt it by December 2024. The analysis window was from November 2023 to December 2024, with pre-adoption months serving as the pre-treatment period.

\subsection{Results}

The adoption of TC resulted in a clear and consistent positive impact on developers’ productivity:
\begin{itemize}
\item Significant increase (17.5\% with 95\% confidence interval [15.9\%, 19.0\%]) in CLT per user-month (see Figure below for more details).
\item Decrease in MeanDurInvSess (-3.6\% [-3.9\%, -3.2\%]).

\item Active Coding time per CL considering CL size is not impacted statistically significantly.

\end{itemize}

\begin{figure}
\includegraphics[scale=0.25]{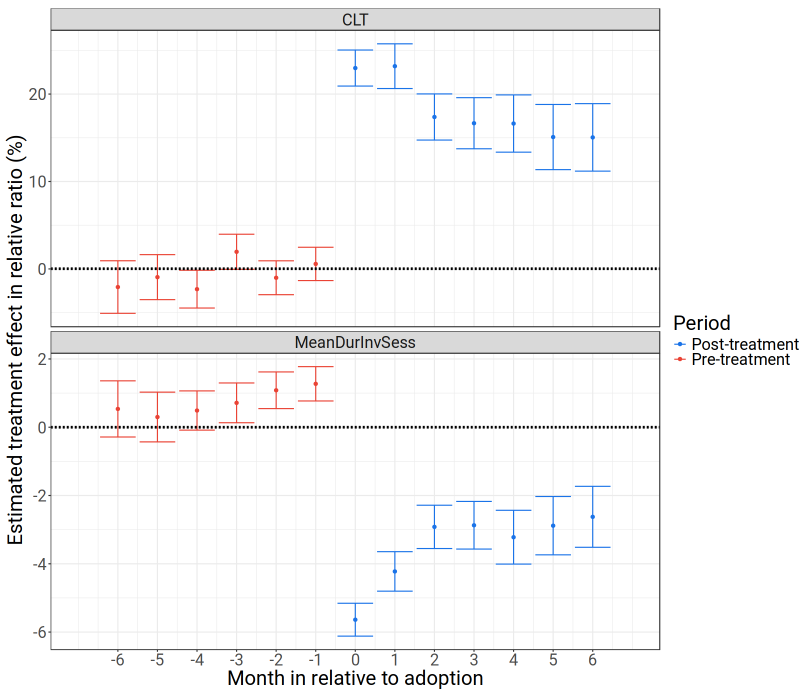}
\caption{Average treatment effect in CLT and MeanDurInvSess across all treatment cohorts before and after adoption \label{fig:prod}}
\end{figure}

Fig~\ref{fig:prod} shows the aggregated treatment effects for CLT and MeanDurInvSess across all cohorts, for months relative to the first adoption; the negative months represent months before adoption, and positive months represent after. We can see that the differences between treatment vs. control are not significant in the pre-treatment months (in red), which shows that the selection bias has been properly mitigated by the methods. These non-significant effects serve as a placebo check, confirming that the treatment and control groups followed parallel trends before the feature was adopted. The treatment effects in the post-treatment months are apparent, with the biggest effects seen in the first two months, and smaller but still positive in the later months. This could be due to the novelty effect and the excitement from adopting a new tool.

\subsection{Robustness and Sensitivity}

The key benefit of this study is its analysis of developer productivity during regular day-to-day work. This differentiates it from many other lab studies in the industry, which analyze productivity based on specific tasks completed with and without AI assistance, while those tasks are very limited in representing an actual developer’s work week \cite{paradis2024doesaiimpactdevelopment,peng2023impactaideveloperproductivity,peng2024airevolutionchatbot}. 

\textbf{Robustness and Alternative Specifications}: To calibrate the strength of our findings, we performed a series of robustness checks. First, as noted in Section 4.3, the lack of significant effects in the pre-treatment period (Fig. 4) serves as a placebo test, supporting the parallel trends assumption of the DiD framework. Second, we verified that the results are robust to alternative specifications. We observed ballpark similar magnitudes of impact when using a fixed-effect model instead of DiD, stratifying the model by CL size and workspace type, and adjusting for different user cohorts—such as excluding those who had adopted other AI tools prior to Transform Code or excluding users with large changes in active coding time. The consistency across these variations reinforces the reported +17.5\% CLT increase.

\textbf{Caveats and Sensitivity}: Despite these checks, our analysis remains based on observational data. While we controlled for many factors such as role, region, and tenure, unobserved confounders like the specific phase of work (e.g., planning vs. active coding) could influence adoption timing. While this sensitivity to unobserved factors means the precise magnitude of the CLT increase cannot be definitively guaranteed, the persistence of the positive effect over multiple months suggests a genuine productivity shift rather than a transient selection effect.  

Our analysis--albeit focused on TC--demonstrates that AI-powered software engineering features can significantly enhance developer productivity internally at Google.

\section{Conclusion}

We believe above discussion will help applied ML teams in the industry working on AI coding products with a holistic approach towards productivity improvements of software engineers, including:

\begin{enumerate}
\item iteration on the user experience (e.g. discoverability, ease of suggestion review)
\item integration (e.g. optimizing for latency, providing relevant context)
\item model capability tuning (e.g. with data from real product usage)
\item measuring tactical improvements with product-specific metrics (e.g. acceptance rate, FCML)
\item validating overall progress by carefully measuring productivity metrics (e.g. throughput)
\end{enumerate}

Importantly, there are inter-dependencies between the items above. To successfully and quickly land an improvement, there is often the need to make changes across multiple layers of the stack, as well as the need for an effective collaboration between people or teams involved.

\section{Outlook}
\label{sec:outlook}

Looking ahead, we see three milestones in the upcoming progress where AI is transforming software engineering in enterprises:
\begin{enumerate}
\item AI acts as a pair programmer accelerating software engineers in some tasks
\item AI fulfills individual tasks from the software development lifecycle (SDLC) under human supervision
\item AI owns full SDLC journeys with human supervision
\end{enumerate}

The features we discuss above are all part of milestone 1. We expect code completion’s scope to expand into assisting with followup-edits beyond the position of the cursor, e.g. for followup fixes of copy-pasting code \cite{ForteRevaj2024,nguyen2025smartpasteautomaticallyfixing}. 

Over the last few quarters, we increasingly see successful early exploratory products (e.g. Cursor, WindSurf, and others) using agentic capabilities towards accomplishing more complex tasks towards milestone 2. However, these products have not been in enterprise usage long enough to permit the depth of insights as the features presented in this article. We do expect significant benefits from fully or partially automating some maintenance tasks that arise in enterprise contexts where maintenance tasks and incremental code changes are as common, if not more, than greenfield development. We previously discussed exploratory efforts around code cleanups, migrations and bug fixing \cite{nikolov2025googleusingaiinternal}. Novel agentic approaches achieving complex tasks in the IDE with human assistance are advancing quickly. Although those products can be considered related to the above discussion of Transform Code, long-running asynchronous agents pose novel challenges on IDE UX. 

An example of a feature from milestone 3 would be to  use production signals (e.g. about resource usage, logs, product metrics) or planning signals (e.g. prioritized product needs),  generate the needed changes in the code base, deploy safely in production (potentially as experiments), collect success signals, and involve the human developers for review where needed.

We expect AI assistance to progress gradually, where human developers are increasingly more empowered to drive development of new features more quickly from idea to products. 

\begin{acks}
The authors would like to thank our collaborators at Google. Most of the work presented here, has been accomplished in collaboration with a large group of contributors working on the coding tools, research and on various other aspects that were critical for the success of the discussed features.
\end{acks}

\bibliographystyle{ACM-Reference-Format}
\bibliography{bibliography} 

\end{document}